\documentclass[pra,showpacs]{revtex4-1}
\usepackage{graphicx}
\usepackage{amsmath}
\usepackage{amssymb}

\begin{document}

\title{Non-separated states from squeezed dark-state polaritons in electromagnetically-induced-transparency media}
\author{You-Lin Chuang$^{1,2}$, Ite A. Yu$^{2}$, and Ray-Kuang Lee$^{1,2,3}$}
\affiliation{
$^{1}$Institute of Photonics Technologies, National Tsing-Hua University, Hsinchu 300, Taiwan\\
$^{2}$Department of Physics, National Tsing-Hua University, Hsinchu 300, Taiwan\\
$^3$ Physics Division, National Center of Theoretical Science, Hsinchu 300, Taiwan}

\begin{abstract}
Within the frame of quantized dark-state polaritons in electromagnetically-induced-transparency media, noise fluctuations in the quadrature components are studied. 
Squeezed state transfer, quantum correlation, and noise entanglement between probe field and atomic polarization are demonstrated  in  single-  and double-$\Lambda$ configurations, respectively.
Even though a larger degree of squeezing parameter in the continuous variable helps to establish stronger quantum correlations, inseparability criterion is satisfied only within a finite range of squeezing parameter.
The results obtained in the present study may be useful for guiding experimental realization of quantum memory devices for possible applications in quantum information and computation.
\end{abstract}


\maketitle

\section{INTRODUCTION}
It is believed that quantum network has a great potential than the classical one  in providing many powerful applications for quantum information science \cite{longdis, npp}. Not only  theoretical schemes  \cite{QC,QR,CVEIT,CQED1,CQED2,CQED3,Opticalqmemory, slowsoliton}, but also experimental implementations \cite{single,EITsingle,CEIT, QSS} are demonstrated in various systems, intending to manipulate and control the quantum objects. 
Among the candidates as  quantum bits, photon, the quanta of light,  is the fastest and robust carrier in the quantum network. 
For the storage and retrieval of optical information, electromagnetically-induced-transparency (EIT) system serves as an ideal quantum interface between photon and atoms \cite{EIT1,EIT2}. 
Based on quantum coherent interference, profile as well as the phase of optical information are well controllable and perfectly preserved in the adiabatic condition \cite{Manipulate1, Manipulate2}. 
Moreover, instead of using classical light source, non-classical states are also investigated in the EIT system, in order to map  quantum state of light  onto atomic ensembles as a quantum memory device~\cite{Manipulate3,Manipulate4,LS1,LS2, ss3}.

Recently, experimental progresses include the slowing-down of squeezed vacuum pulse~\cite{ss1, ss2} and the storage of squeezed states for several micro seconds~\cite{st1,st2,st3}.
Since the photon statistics of squeezed light differs from the Poisson distribution, a full quantum theory for  the storage and retrieval of non-classical light is needed.              
Based on the perturbed quantum fluctuations, for quasi-continuous wave inputs, EIT media  become opaque for squeezed states, with  an oscillatory transfer of the initial quantum properties between the probe and pump fields ~\cite{Barberis}.
The entanglement in quantum fluctuation of electromagnetic fields is possible to be preserved or to be produced through an EIT medium \cite{entangleEIT}.
Furthermore, through the picture of dark-state polaritons,  quantum state transfer between optical pulse  and atomic polarization  is clearly illustrated during the storage and retrieval process  \cite{SQtransfer, DSP1, DSP2}.

In addition to the quantum state transfer, in this work, we introduce squeezed dark-state polaritons by the corresponding squeezed operator, and study the  quantum correlation and entanglement of noise fluctuations in the quadrature components during the storage and retrieval process. 
As one may expect, when the squeezing parameter  $\text{r} =0 $ (a coherent state), there is no quantum correlation between probe field and atomic polarization; while a larger degree of squeezing parameter, the stronger quantum correlation is established.
In the contrary, inseparability criterion to guarantee an entanglement state is satisfied only with a finite range of squeezing parameter.
Extension to a double-$\Lambda$ configuration is also studied, in orde to reveal the conditions to have mutual entanglement among the noise correlations of two probe fields, and one common atomic polarization. Our results pave the way to implement the quantum interface between photon and atomic system.

The remaining part of this paper is organized as follows. In Sec. $\textrm{II}$, we start from the picture of quantized dark-state polaritons, and derive related quadrature variance in the noise fluctuations for the field and atomic operators.
Quantum correlation and entanglement between field and atomic polarization operators during storage and retrieval process is demonstrated. Especially, in Sec. $\textrm{III}$, we address the  inseparability  condition in the continuous variables for the quadrature components of field and atomic operators in a single-$ \Lambda$ configuration. 
The generalization to a double-$\Lambda$ configure is extended in 
Sec. $\textrm{IV}$, where the quantum variances of two quantized probe fields and atomic polarization are shown.
Finally, we give a brief conclusion in Sec. $\textrm{V}$.

\begin{figure}
\includegraphics[width=8.0cm]{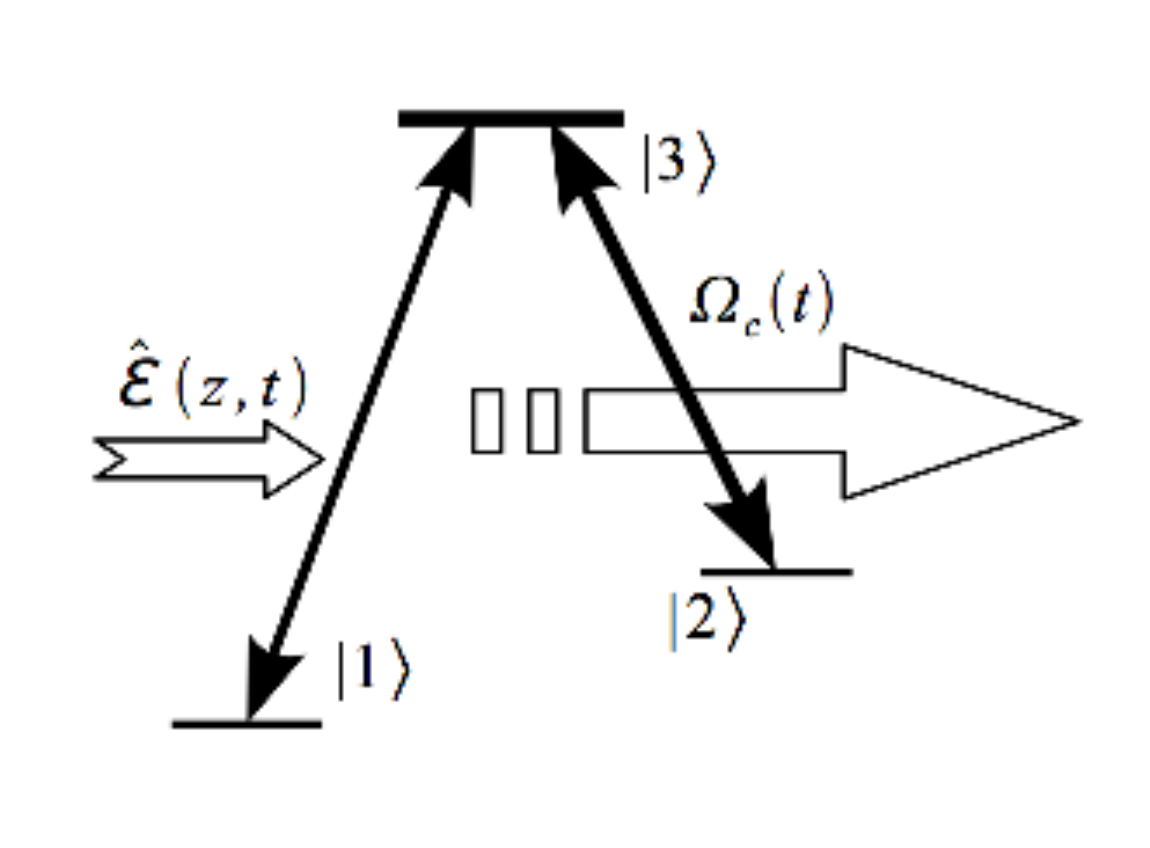} 
\caption{The EIT system considered in a single-$\Lambda$ configuration, where the transitions $\vert 1 \rangle \leftrightarrow  \vert 3\rangle$ and $\vert 2 \rangle \leftrightarrow  \vert 3\rangle$ are driven resonantly by a
quantized probe field, $\hat{\mathcal{E}}$, and a classical coupling field, denoted by its Rabi frequency $\Omega_c$, separately.}
\end{figure}

\section{Squeezed dark-state polaritons in a single-$\Lambda$ configuration}
We begin with  the EIT system in a single-$ \Lambda $ configuration, as illustrated in Fig. 1.
Here, two co-propagating beams pass through a three-level atomic ensemble in the $z$ direction, with the total number of atoms denoted by $N$.
The probe field excites the transition from the state $ \vert 1\rangle$ to the state $\vert 3\rangle $, which is treated by the quantum  field operator $ \hat{\mathcal{E}} (z,t)$ in the slowly varying envelope approximation. 
The transition between $ \vert 2\rangle$ and $\vert3\rangle $  is driven resonantly by a classical coupling field with the Rabi frequency denoted by $\Omega_c(t)$, which is a time-dependent function during the storage and retrieval process. 
In the Heisenberg picture, the interaction Hamiltonian for such a single-$\Lambda$ EIT system is  given as~\cite{EIT1, EIT2}
\begin{eqnarray}
\hat{H} = -\left( \hbar\, g\,\hat{\sigma}_{31}\,\hat{\mathcal{E}} + \hbar\, \Omega_c\, \hat{\sigma}_{32} + {\text \cal H.C.}\right), 
\end{eqnarray} 
where  $\hat{\sigma}_{\mu\nu} = \vert \mu \rangle\langle \nu \vert $ $(\mu, \nu = 1, 2, 3)$ is used as the collective atomic operator, the atom-field coupling strength for the transition $\vert 1\rangle \leftrightarrow  \vert3\rangle$ is denoted by the constant $g$, and ${\text \cal H.C.}$ represents the Hermitian conjugate,

It is known that with the low-intensity approximation, $\langle\hat{\sigma}_{11}\rangle \approx 1$,
 and adiabatic limit, $\hat{\sigma}_{12} \approx -\dfrac{g}{\Omega_c}\hat{\mathcal{E}}$, the propagation of quantum fields in EIT media can be described by the  dark-state polariton~\cite{DSP1, DSP2}, 
\begin{eqnarray}
\hat{\Psi}(z,t) = \cos \theta(t)\, \hat{\mathcal{E}}(z,t) - \sqrt{N}\sin\theta(t)\, \hat{\sigma},
\end{eqnarray}
which is a linear superposition of field and atomic operators.
Here, $\hat{\sigma} \equiv \hat{\sigma}_{12} $ is used for the atomic polarization between two lower states, $\vert 1\rangle$ and $\vert 2\rangle$.
In general, the rules of  commutation relation for bosonic fields $\hat{\mathcal{E}}$ and atomic polarization $\hat{\sigma}$ are different, {\it i.e.},
\begin{eqnarray}
&&\left[ \hat{\mathcal{E}} (z, t),\hat{\mathcal{E}}^{\dagger}(z, t_1)\right] = T\,\delta(t-t_1),\\
&&\left[ \hat{\sigma} (z, t),\hat{\sigma}^{\dagger}(z, t_1)\right] = - \dfrac{\hat{\sigma}_{22}-\hat{\sigma}_{11}}{N} \,T\,\delta(t-t_1).
\end{eqnarray}
where $T$ is the characteristic time scale.
Here, we have applied the {\it equal space} commutation relations and a single longitudinal mode for the field and atomic systems is used, too.
However, if we assume that the atomic system is originally in the ground state,  $\langle\hat{\sigma}_{22}-\hat{\sigma}_{11}\rangle \approx -1$~\cite{SIT}, this dark-state polariton $\hat{\Psi}(z,t)$ is a quasi-particle satisfying the Bosonic commutation relation:
\begin{eqnarray}
&&\left[ \hat{\Psi} (z, t),\hat{\Psi}^{\dagger}(z, t_1)\right] \approx T\,\delta(t-t_1),
\end{eqnarray} 
where  the characteristic time scale $T$ can be obtained by requiring $T^{-1}\int_0^T\left[ \hat{\Psi} (z, t),\hat{\Psi}^{\dagger}(z, t_1)\right]\,dt = 1$.
However, the commutation relations between the dark-state polariton and field (atomic) operators are
\begin{eqnarray}
&&\left[ \hat{\Psi} (z, t), \hat{\mathcal{E}}^{\dagger}(z^\prime, t_1)\right] = \cos \theta(t)\,T\,\delta(t-t_1),\\
&&\left[ \hat{\Psi} (z, t), \hat{\sigma}^{\dagger}(z^\prime, t_1)\right] = -\dfrac{\sin \theta(t)}{\sqrt{N}}\, T\, \delta(t-t_1).  
\end{eqnarray} 
Under the picture of dark-state polaritons, the governing equation of motion during the storage and retrieval process is 
\begin{eqnarray}
\left[ \dfrac{\partial}{\partial t}+v_g(t) \dfrac{\partial}{\partial z}\right]\hat{\Psi}(z,t) = 0, \label{DSP1}
\end{eqnarray}
where  the group velocity  of dark-state polartion is given by $v_g(t) = c \cos^2\theta(t) $, with the speed of light in the vacuum $c$, and $\theta(t) = \tan^{-1}[g\sqrt{N}/\Omega_c(t)]$ accounts the mixing angle as a function of time.
Without any decay mechanism, the evolution of a dark-state polariton is described by changing the value of $\Omega_c$(t) with respect to the time.
When $\theta(t) = 0$, or $\Omega_c(t)/g \rightarrow \infty$, the dark-state polariton is said to be a {\it photon-like} state, {\it i.e.}, $\hat{\Psi} = \hat{\mathcal{E}}$; while $\theta(t) = \pi/2$, or $\Omega_c(t)/g = 0$, the dark-state polariton is a {\it atom-like} state, {\it i.e.}, $\hat{\Psi} = -\sqrt{N}\hat{\sigma}$.

Based on the quantized polariton field operator, $\hat{\Psi}(z, t)$, in the following we introduce the squeezed state for dark-state polaritons by defining a squeezing operator $\hat{S}(\xi)$: 
\begin{eqnarray}
\hat{S}(\xi) = \exp\left( \dfrac{\xi^{\ast}}{2} \int_0^T \hat{\Psi}^2\,dt -  \dfrac{\xi}{2}\int_0^T \hat{\Psi}^{\dagger2} \, dt\right), 
\label{SO}
\end{eqnarray}
where $\xi = \text{r} e^{i\delta} $ denotes the degree of noise squeezing, with the squeezing parameter  $\text{r} = \vert\xi\vert $ and the related squeezing angle $\delta$.
The corresponding squeezed  vacuum state for a dark-state polariton is represented in the basis of  $\vert\xi\rangle = \hat{S}(\xi)\vert 0\rangle $, which is composited by the vacuum state of fields and ground state of atomic polarization, {\it i.e.}, that is $ \vert 0\rangle = \vert 0\rangle_{\text{field}}\otimes \vert 1 \rangle_{\text{atom}}$.
For the quantum noises in continuous variables, we have the related quadrature operator as $\hat{X}_{\Psi} = \hat{\Psi}+\hat{\Psi}^{\dagger}$ for the amplitude (in-phase) fluctuations.
With above definitions, the quadrature variance of dark-state polaritons is found to be
\begin{eqnarray}
\Delta X_{\Psi}^2 &\equiv& \langle \xi \vert \Delta \hat{X}_\Psi^2\vert\xi\rangle, \nonumber\\
&=& \cos^2\theta(t)\Delta X_{\mathcal{E}}^2 + N \sin^2\theta(t) \Delta X_{\sigma}^2 \\\nonumber
&-&\sqrt{N} \sin\theta(t)\cos\theta(t)\left[ \langle \hat{X}_{\mathcal{E}}\hat{X}_{\sigma}\rangle + \langle \hat{X}_{\sigma}\hat{X}_{\mathcal{E}}\rangle\right].  
\label{Xdsp}
\end{eqnarray} 
Here, the first and second terms, {\it i.e.}, $\Delta X_{\mathcal{E}}^2 
\equiv \langle \xi \vert \Delta \hat{X}_\mathcal{E}^2\vert\xi\rangle$ and $\Delta X_{\sigma}^2
\equiv \langle \xi \vert \Delta \hat{X}_\sigma^2\vert\xi\rangle$,  are the corresponding quadrature variances of field and atomic parts, with the  in-phase quadrature components $\hat{X}_{\mathcal{E}} = \hat{\mathcal{E}}+\hat{\mathcal{E}}^{\dagger}$ and $\hat{X}_{\sigma} = \hat{\sigma}+\hat{\sigma}^{\dagger}$ defined for the field and atomic operators, respectively.
It can be seen that for a dark-state polariton, the quadrature variances of  filed and atomic operators are added together by the time-dependent coefficient $\theta(t)$, or $\Omega_c(t)$, during the storage and retrieval process.
Furthermore,  we also have contributions from the correlation between the field and atomic ensemble, {\it i.e.}, $\langle \hat{X}_{\mathcal{E}}\hat{X}_{\sigma}\rangle$ and $\langle \hat{X}_{\sigma}\hat{X}_{\mathcal{E}}\rangle$.

Consider possible experimental demonstration, one can use a squeezed light source for the probe field.
For a given initial quadrature variance, $\Delta X_{\Psi}^2(t= 0) \equiv \Delta X_{\text{in}}^2$, 
 we can manipulate the distribution of quantum noise fluctuations  between the field and atomic parts, that is 
\begin{eqnarray}
&&\Delta X_{\mathcal{E}}^2 = \dfrac{(\dfrac{\Omega_c}{g})^2 (\Delta X_{\text{in}}^2)+ N}{(\dfrac{\Omega_c}{g})^2 + N}, \\
&&\Delta X_{\sigma}^2 = \dfrac{1}{N}\left[\dfrac{(\dfrac{\Omega_c}{g})^2 +  N(\Delta X_{\text{in}}^2)}{(\dfrac{\Omega_c}{g})^2 + N}\right].
\end{eqnarray}
One can see that in the limit $\Omega_c/g \rightarrow \infty$, we have $\Delta X_{\mathcal{E}}^2 = \Delta X_{\text{in}}^2$ and $\Delta X_{\mathcal{\sigma}}^2 = 1/N$ for a photon-like dark-state polariton. 
In the other limit $\Omega_c/g = 0$, the noise fluctuations for an atom-like dark-state polariton are $\Delta X_{\mathcal{E}}^2 = 1$ and $\Delta X_{\mathcal{\sigma}}^2 = \Delta X_{\text{in}}^2/N$.

Consider typical  experimental conditions in the realization of EIT phenomena, such as the systems of cold  $^{87}$Rb atoms, we have $(\Omega_c/g)^2 \ll N$. In this scenario, since the dark-state polartion is in the atom-like state,  the quantum fluctuation of a dark-state polariton is dominated by the atomic quadrature variance.     
With this condition, the quadrature variances can be approximated by
\begin{eqnarray}
&&\Delta X_{\mathcal{E}}^2 \simeq 1- \dfrac{1}{N}(\dfrac{\Omega_c}{g})^2(1-\Delta X_{\text{in}}^2), \\
&&\Delta X_{\sigma}^2 \simeq \dfrac{1}{N}\left[ \Delta X_{\text{in}}^2+  \dfrac{1}{N}(\dfrac{\Omega_c}{g})^2(1-\Delta X_{\text{in}}^2)\right].  
\end{eqnarray}
We want to remark that when the initial quadrature of input state is a coherent state, {\it i.e.}, $\Delta X_{\text{in}}^2 = 1$, the noise variance in the field component remains as the same as  that of  vacuum states; while the quantum fluctuation in the  atomic component corresponds to that of a spin coherent state. 
That is, when a coherent state is used as the input (a classical light source), both the quantum noise variance in the field and atomic components are independent from the value of control field, $\Omega_c(t)$.

\begin{figure}
\includegraphics[width=8.0cm]{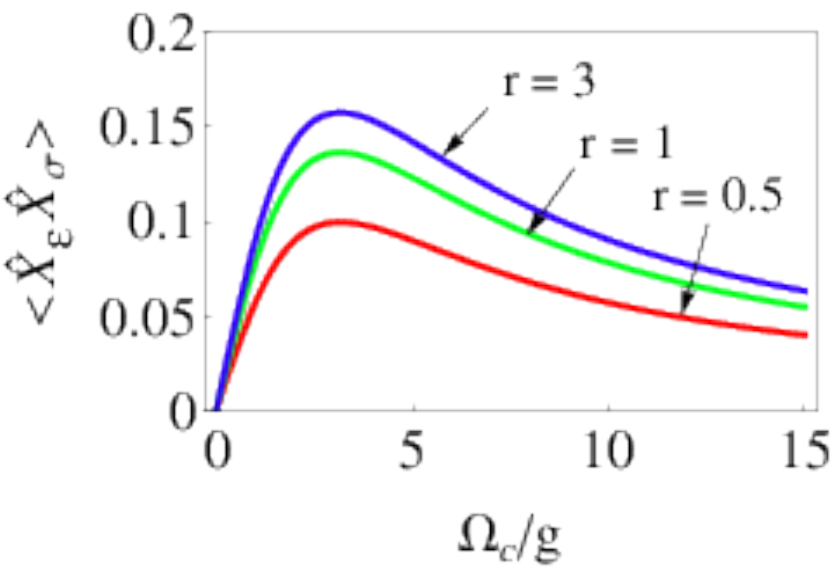} 
\caption{(Color online) Quantum correlation between the field and atomic polarization, $\langle \hat{X}_{\mathcal{E}}\hat{X}_{\sigma}\rangle$, shown as a function of the normalized control field, $\Omega_c/g$, for different values of squeezing parameter, $\text{r}$. Here, the number of atoms is fixed at $N=10$.}
\end{figure}

Naively, one may take an EIT media as a linear system, and expect a complete transfer for  the non-classical properties from input field to the atomic system under the picture of dark-state polaritons.
Nevertheless, due to the last term in Eq. (10), non-trivial quantum correlations between the field and atomic operators will be shown  through the quantum noise squeezing. 
Here, the quantum correlation between field and atomic components of a dark-state polartion has the form:
\begin{eqnarray}
\langle \hat{X}_{\mathcal{E}}\hat{X}_{\sigma}\rangle = \langle \hat{X}_{\sigma}\hat{X}_{\mathcal{E}}\rangle = \dfrac{\Omega_c/g}{(\Omega_c/g)^2+N}(1-\Delta X_{\text{in}}^2).
\end{eqnarray}
Again, for an initial coherent state, $\Delta X_{\text{in}}^2 = 1$, the quantum correlation between field and atomic operators is zero. 
In Fig. 2, we show the quantum correlation between the field and atomic polarization, $\langle \hat{X}_{\mathcal{E}}\hat{X}_{\sigma}\rangle$, as a function of the normalized control field, $\Omega_c/g$, for different values of squeezing parameter, $\text{r}$.
As expected, the more non-classical properties, with a larger value of the squeezing parameter $\text{r}$, the stronger quantum correlation is. 
Moreover, when the dark-state polariton can be approximated by a photon-like state ($\Omega_c/g \rightarrow \infty$) or an atom-like state ($\Omega_c/g = 0$), the quantum correlation becomes zero as well. 
That is, the quantum correlation exists only with a superposition of partial-photon and partial-atom states.

\begin{figure}
\includegraphics[width=8cm]{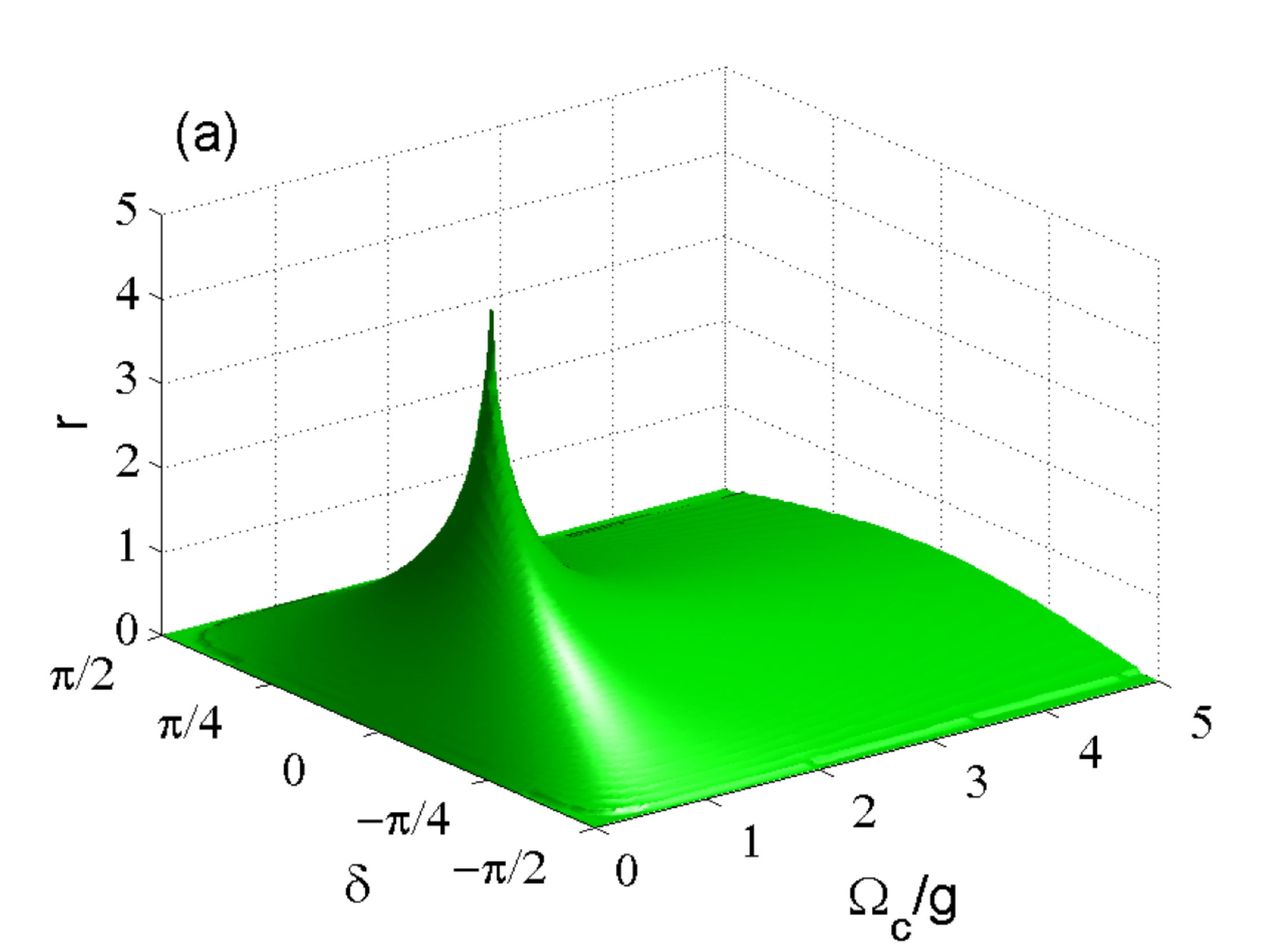}
\includegraphics[width=8cm]{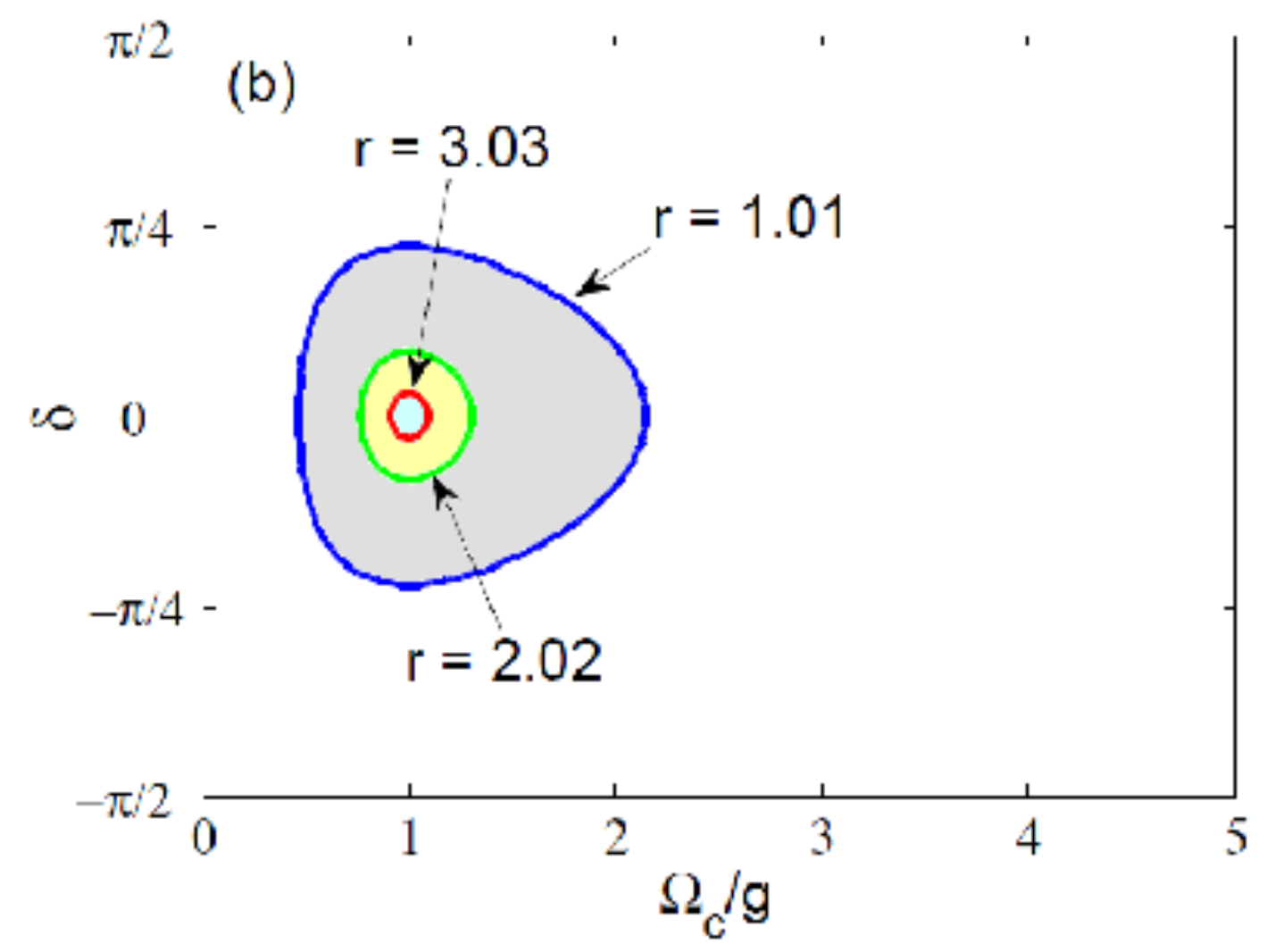}
\includegraphics[width=8cm]{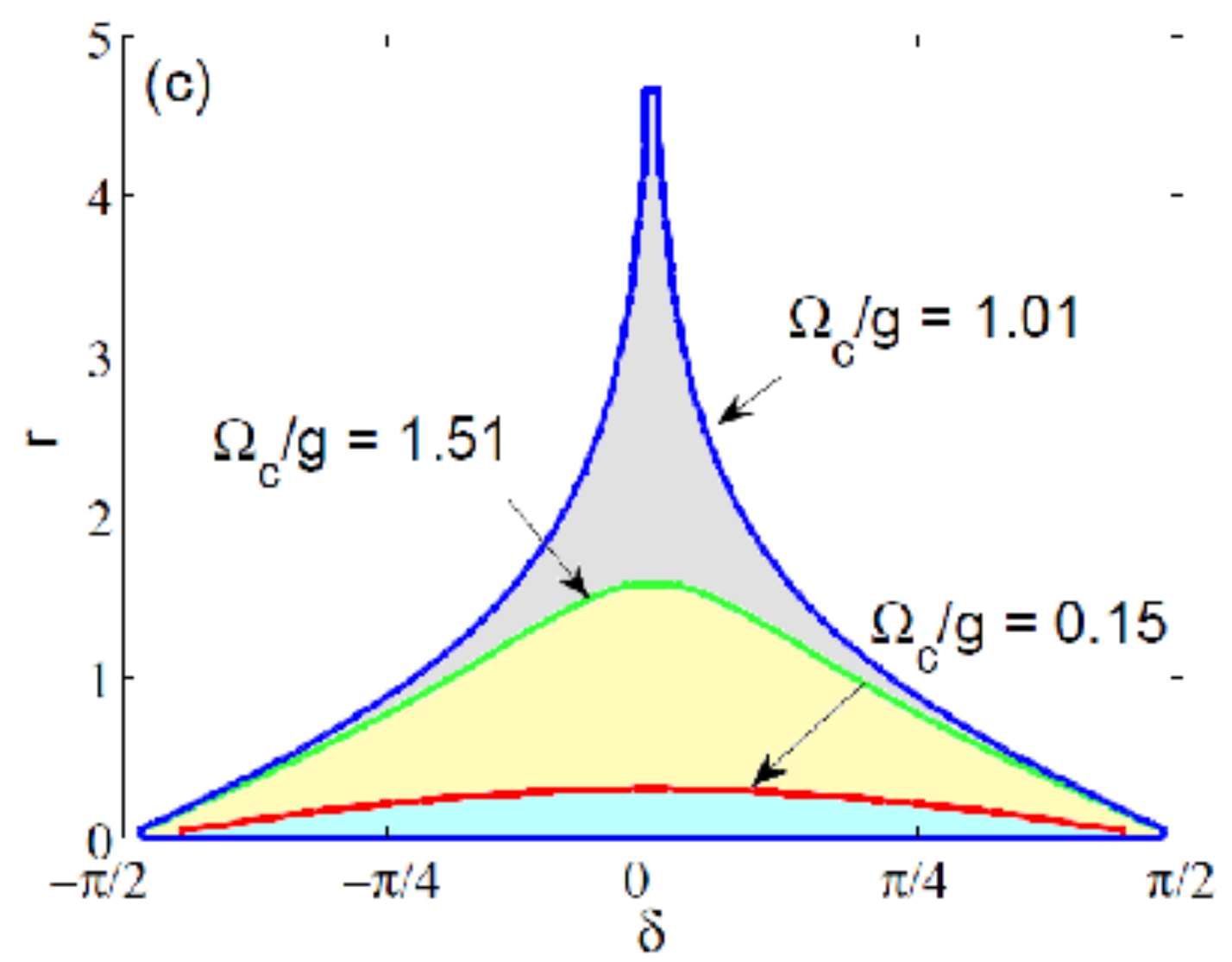}
\includegraphics[width=8cm]{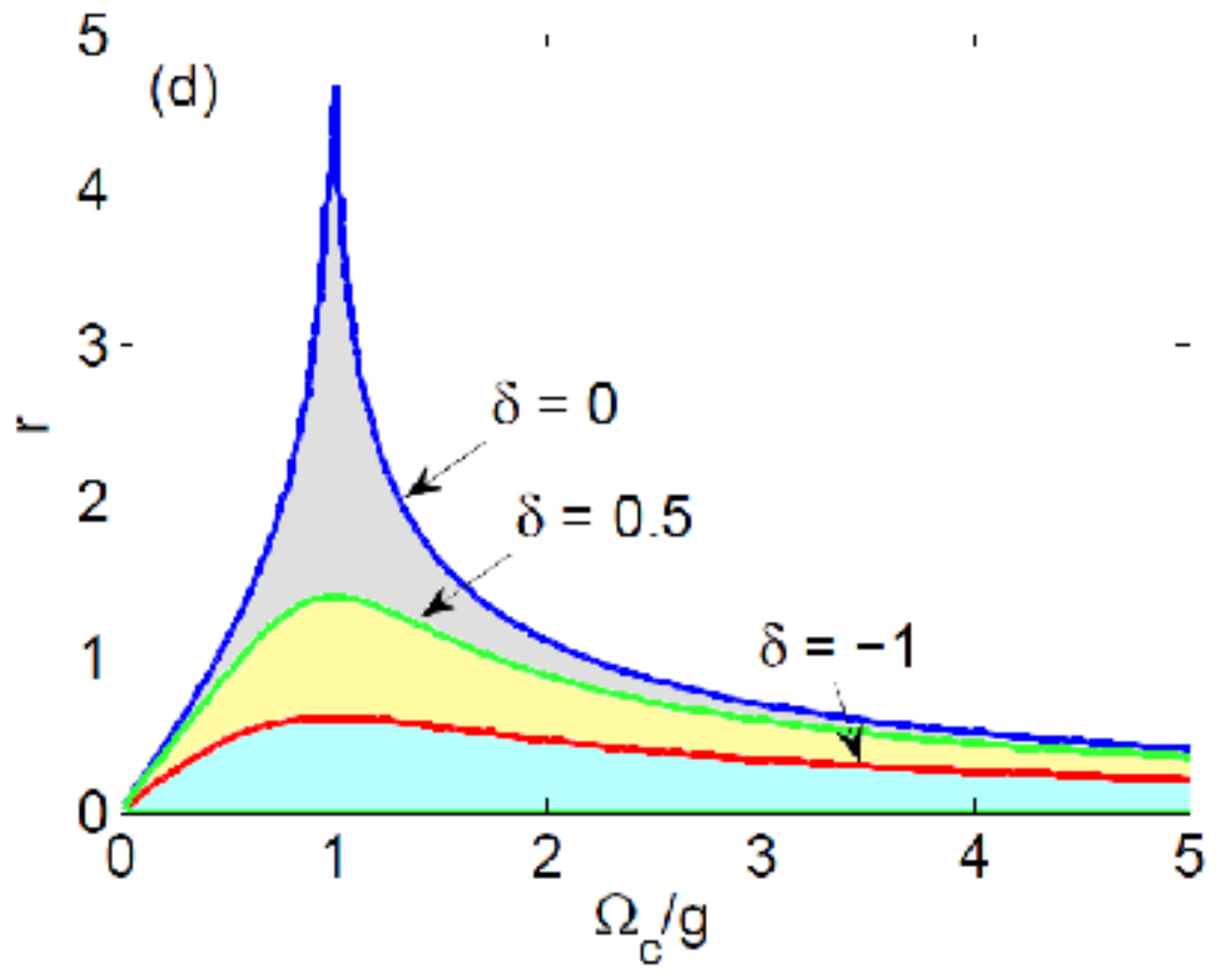}
\caption{(Color online) (a) The surface for inseparability condition, defined by requiring the function $\mathcal{F}(\Omega_c/g, \text{r}, \delta) = 0$ shown in Eq. (18). The parameter space is expanded by the normalized control field $\Omega_c/g$, the degree of squeezing parameter $\text{r}$, and the squeezing angle $\delta$. The contour plots are obtained by projecting the surface into the plane of  (b) $(\Omega_c/g, \delta)$, (c) $(\delta, \text{r})$,  and (d) $(\Omega_c/g, \text{r})$, respectively, with the other parameter shown in the markers. The colored regions indicate the parameter space that $\mathcal{F}(\Omega_c/g, \text{r}, \delta) < 0$.}
\end{figure}  

\section{Inseparability condition for squeezed dark-state polaritons}
Next, we study the entanglement between the quadrature components of field and atomic ensemble, by using the inseparability criterion  for bipartite continuous variables ~\cite{ent, RKL}. Only when the following inequality is satisfied, a bipartite system is said to be entangled, 
\begin{eqnarray}
I_c \equiv\Delta (\hat{X}_\mathcal{E} - \hat{X}_{\sigma})^2 + \Delta (\hat{Y}_\mathcal{E} + \hat{Y}_{\sigma})^2 < 2 + \dfrac{2}{N},
\label{encond}
\end{eqnarray}
where $\hat{X}_\mathcal{E}$ and $\hat{X}_\sigma$ correspond to the in-phase quadrature components; while $\hat{Y}_\mathcal{E} \equiv -i(\hat{\mathcal{E}}-\hat{\mathcal{E}}^{\dagger}) $ and $ \hat{Y}_{\sigma} \equiv -i(\hat{\sigma}-\hat{\sigma}^{\dagger}) $ are the out-of-phase quadrature operators for  field and atomic operators, respectively. 
To have a clear comparison, we normalize this inseparability criterion, {\it i.e.},
\begin{eqnarray}
\mathcal{I}_c  &\equiv& \frac{I_c}{2+\dfrac{2}{N}}
\equiv 1 + \dfrac{2}{1+\dfrac{1}{N}}\left[\dfrac{\mathcal{F}(\Omega_c/g, \text{r}, \delta)}{\dfrac{\Omega_c^2}{g^2}+N}\right]\\
&<& 1,\nonumber
\end{eqnarray}
where the numerator in the bracket of Eq. (17) is defined as
\begin{eqnarray}
\mathcal{F}(\Omega_c/g, \text{r}, \delta) 
\equiv (\dfrac{\Omega_c^2}{g^2} + 1)\sinh^2 
\text{r}- 2\dfrac{\Omega_c}{g} \sinh \text{r} \cosh \text{r}\cos\delta.\nonumber\\
\label{FC}
\end{eqnarray}
The non-separation condition is guaranteed only when  $\mathcal{F}(\Omega_c/g, \text{r}, \delta)$ is smaller than $0$.
According to the inseparability condition shown in Eq. (18), it is obvious that  for a coherent state at the input,  $\text{r}=0$, we do not have non-separated states, {\it i.e.}, $\mathcal{F} = 0$, no matter what the value of the control field $\Omega_c(t)$ is.    
Moreover, the existence of entanglement is independent from the number of atoms, $N$,  despite that the value of $\mathcal{I}_c$ ($I_c$) changes with the number of atoms.

In order to demonstrate the inseparability condition, in Fig. 3(a), we show the surface obtained by requiring the function $\mathcal{F}(\Omega_c/g, \text{r}, \delta)$ in Eq. (18) to be zero,  which gives the border between separated and non-separated states.
Here, the parameter space is expanded by the normalized control field $\Omega_c/g$, the degree of squeezing parameter $\text{r}$, and the related squeezing angle $\delta$. Only the colored region, beneath the surface but above the plane $\text{r} = 0$, supports the non-separated states from squeezed dark-state polaritons during the storage and retrieval process. 

To give a clear illustration, we project the parameter space satisfying  the inseparability condition into the planes of $(\Omega_c/g, \delta)$, $(\delta, \text{r})$, and $(\Omega_c/g, \text{r})$ in Fig. 3(b), 3(c), and 3(d), respectively.
As the same scenario in the quantum correlation between the field and atomic polarization shown in Fig. 2,  it can be seen in Fig. 3(b) that  the non-separated states (the colored regions) are also not supported when $\Omega_c/g = 0$ or $\Omega_c/g \rightarrow \infty$.
Moreover, these non-separated states are measured dominantly along the angle $\delta = 0$, as shown in Fig. 3(c), due to the reason that we have assumed the phase difference between the field and atomic operators is zero.
However, as shown in Fig. 3(d),  the non-separated states are only supported within a finite range of squeezing parameter, $\text{r}$. 
Counter-intuitively, for a larger degree of squeezing parameter, which is believed to possess more non-classical properties, the corresponding inseparability criterion happens to be invalid.

\begin{figure}
\includegraphics[width=8cm]{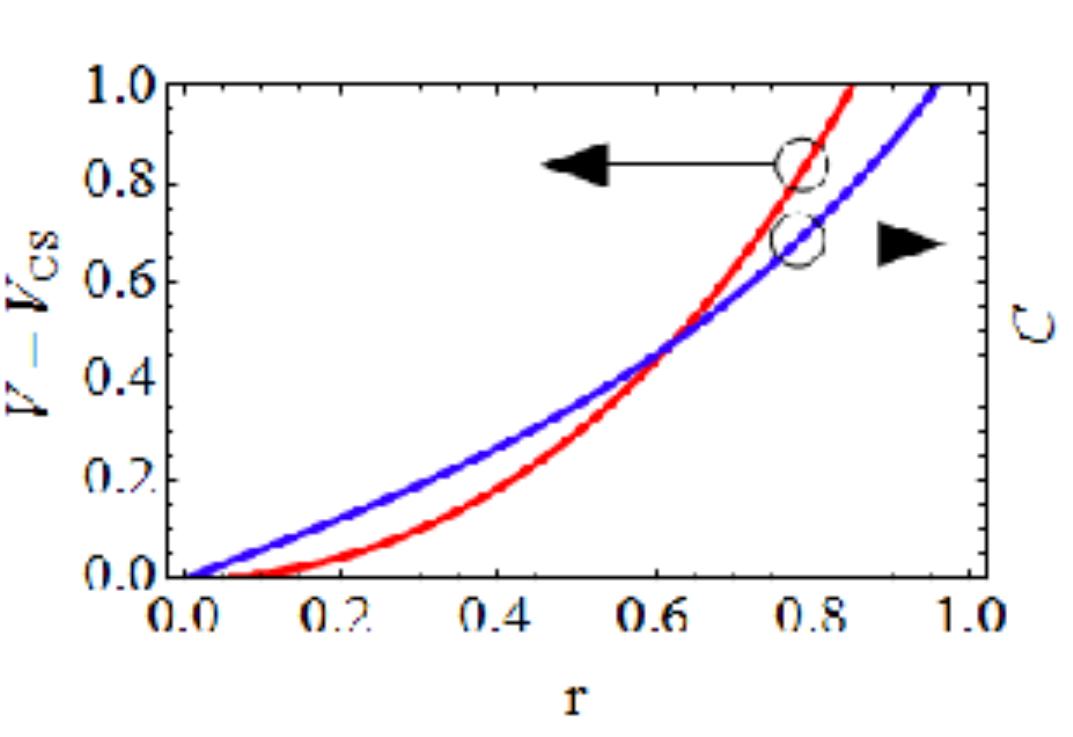} 
\caption{(Color online) Competition between the terms $(V-V_\text{SC})$ and $C$ in the inseparability criterion, shown in Eq. (19). Here,  $V= \Delta X_{\mathcal{E}}^2 + \Delta Y_{\mathcal{E}}^2 + \Delta X_{\sigma}^2 + \Delta Y_{\sigma}^2$ and $V_{CS} = 2+2/N$ are the sum of quadrature variances in the bipartite system and coherent states, respectively; while $C = 2\langle\hat{X}_{\mathcal{E}}\hat{X}_{\sigma}\rangle - 2\langle\hat{Y}_{\mathcal{E}}\hat{Y}_{\sigma}\rangle$ denotes the difference in the quantum correlations between two orthogonal quadratures. Other parameters used are $\Omega_c/g = 0.3$ and $\delta = 0$.} 
\end{figure} 

The reason why only a finite range of squeezing parameter supports non-separated states can be illustrated in the following way.  In terms of the quadrature variances, the inseparability criterion in Eq. (16) can be re-written as
\begin{eqnarray}
&& I_c  - (2+\dfrac{2}{N}) \nonumber\\
&&= [\Delta X_{\mathcal{E}}^2 + \Delta Y_{\mathcal{E}}^2 + \Delta X_{\sigma}^2 + \Delta Y_{\sigma}^2] - [(2+\dfrac{2}{N})]\nonumber\\
 &&- 2[ \langle\hat{X}_{\mathcal{E}}\hat{X}_{\sigma}\rangle - \langle\hat{Y}_{\mathcal{E}}\hat{Y}_{\sigma}\rangle]\nonumber\\
&&\equiv V - V_{\text{CS}}- C < 0.
\end{eqnarray}
From above expansion, It can be seen that to satisfy the inseparability criterion, we have competing terms in Eq. (19). They corresponds to the sum of total variances of field and atomic fluctuations both in the in-phase and out-of-phase quadratures, $V \equiv \Delta X_{\mathcal{E}}^2 + \Delta Y_{\mathcal{E}}^2 + \Delta X_{\sigma}^2 + \Delta Y_{\sigma}^2$, the sum of variance for the coherent photon and coherent atomic states, $V_\text{CS} \equiv  2+2/N$, 
and the difference in the quantum correlations between them in two orthogonal quadratures, $C \equiv  2\langle\hat{X}_{\mathcal{E}}\hat{X}_{\sigma}\rangle - 2\langle\hat{Y}_{\mathcal{E}}\hat{Y}_{\sigma}\rangle$.
For a coherent state, the last term is zero, $C = 0$, for there is no quantum correlation existed. As a result, we do not have non-separated states with an input of coherent states. 
Nevertheless, a non-classical state can not always ensure the inseparability. 
In Fig. 4, we plot the curves for  $V-V_\text{CS}$ and $C$, as a function of the squeezing parameter, $\text{r}$. From Fig. 4, we can see that  the entanglement can only happen when the quantum correlations between field and atomic fluctuations  are stronger than the total sum of quadrature variances, {\it i.e.} $C > (V - V_\text{CS})$.

\section{Squeezed dark-state polaritons in a double-$\Lambda$ configuration}
In the single-$\Lambda$ configuration discussed above, quadrature fluctuations between the field and atomic parts can be entangled within some parameter space. 
However, in practical experimental setup, one  may need to measure both the quantum noise fluctuations of probe field as well as the variance of atomic ensemble, via homodyne detection schemes. 
Due to the difficulties in measuring the collective atomic operators, here, we extend the concept of dark-state polaritons from a single-$\Lambda$ configuration to a double-$\Lambda$ one, in order to have possible experimental realizations with the output fields arriving at a detection apparatus.
As illustrated in Fig. 5,  now we have two 
quantized probe fields, $\hat{\mathcal{E}}_1$ and $\hat{\mathcal{E}}_2$ driving resonantly 
to the transitions $\vert 1 \rangle \leftrightarrow  \vert 3\rangle$ and $\vert 1 \rangle \leftrightarrow  \vert 4\rangle$, with the corresponding coupling strengths $g_1$ and $g_2$, respectively.
At the same time,  two  classical coupling fields, denoted by its Rabi frequency $\Omega_1(t)$ and $\Omega_2(t)$ drive the transitions $\vert 2 \rangle \leftrightarrow  \vert 4\rangle$ and $\vert 2 \rangle \leftrightarrow  \vert 3\rangle$, simultaneously.

\begin{figure}
\includegraphics[width=8.0cm]{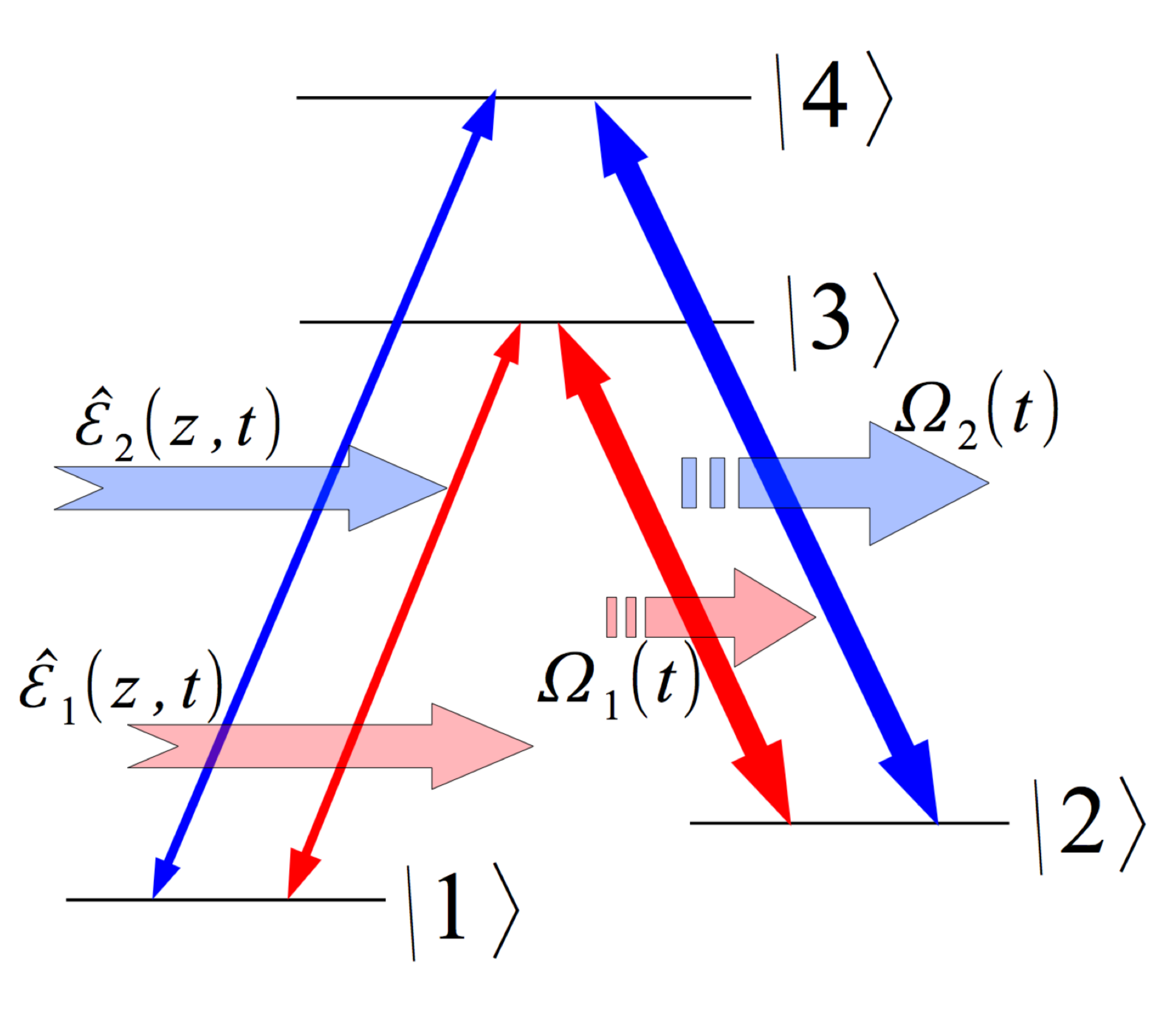} 
\caption{The EIT system considered in a double-$\Lambda$ configuration, where the transitions $\vert 1 \rangle \leftrightarrow  \vert 3\rangle$ and $\vert 1 \rangle \leftrightarrow  \vert 4\rangle$ are driven resonantly by two
quantized probe fields, $\hat{\mathcal{E}}_1$ and $\hat{\mathcal{E}}_2$; while two  classical coupling fields, denoted by its Rabi frequency $\Omega_1$ and $\Omega_2$ drive the transitions $\vert 2 \rangle \leftrightarrow  \vert 3\rangle$ and $\vert 2 \rangle \leftrightarrow  \vert 4\rangle$, simultaneously.}
\label{f3}
\end{figure} 

Due to the share of a common atomic polarization, $\hat{\sigma}_{12}$, we can extend the concept of dark-state polaritons to describe the storage and retrieval process in such a double-$\Lambda$ configuration~\cite{DSP3, DSP4}. 
In this picture, the corresponding quantized dark-state polariton, $\hat{\Psi}$, is composited by two probe field operators, $\hat{\mathcal{E}}_1$ and $\hat{\mathcal{E}}_2$,  and the atomic polarization operator, $\hat{\sigma} \equiv \hat{\sigma}_{12}$, {\it i.e.},
\begin{eqnarray}
\hat{\Psi} = \cos \theta \cos \phi \,\hat{\mathcal{E}}_1 + \cos \theta \sin \phi\, \hat{\mathcal{E}}_2 - \sqrt{N}\sin\theta\,\hat{\sigma},
\label{DSP3}
\end{eqnarray}   
with the mixing angles $\theta(t)$, between the field and atomic polarization, and $\phi(t)$, between two probe fields defined as
\begin{eqnarray*}
\theta(t)  &\equiv& \tan^{-1} \left[ \dfrac{g_1 \sqrt{N}}{\Omega_1}\left( 1 + \dfrac{g_1^2\Omega_2^2}{g_2^2\Omega_1^2}\right)^{-1/2} \right],\\
\phi(t) &\equiv& \tan^{-1} \left( g_1\Omega_2 / g_2\Omega_1\right).
\end{eqnarray*}
 
With the same concept for the squeezed operator introduced in Eq. (9),  the corresponding squeezed state of dark-state polaritons in a double-$\Lambda$ configuration is defined as
$\vert\xi\rangle \equiv \hat{S} \vert 0 \rangle \equiv \vert 0\rangle_{\mathcal{E}_1} \otimes \vert 0\rangle_{\mathcal{E}_2} \otimes\vert 1\rangle_{\text{atom}}$. 
The quadrature variance of this dark-state polariton is found to be 
\begin{eqnarray}
\Delta X_{\Psi}^2 &=& N \sin^2\theta(t) \Delta X_{\sigma}^2 \\
&+&  \cos^2\theta(t)\cos^2\phi(t)\Delta X_1^2 + \cos^2\theta(t)\sin^2\phi(t)\Delta X_2^2 \nonumber\\
&+&\cos\theta(t)\cos\phi(t)\sin\phi(t)[\langle\hat{X}_1\hat{X}_2\rangle + \langle\hat{X}_2\hat{X}_1\rangle]\nonumber\\
&-&\sqrt{N} \sin\theta(t)\cos\theta(t)\cos\phi(t)[\langle\hat{X}_1\hat{X}_{\sigma}\rangle + \langle\hat{X}_{\sigma}\hat{X}_1\rangle]\nonumber\\
&-&\sqrt{N} \sin\theta(t)\cos\theta(t)\sin\phi(t)[\langle\hat{X}_2\hat{X}_{\sigma}\rangle + \langle\hat{X}_{\sigma}\hat{X}_2\rangle]\nonumber.  
\label{Xdsp}
\end{eqnarray} 
where $\hat{X}_i \equiv \hat{\mathcal{E}}_i + \hat{\mathcal{E}}^\dagger_i$, $i =1, 2$, denotes the in-phase quadrature component of probe field, $\hat{\mathcal{E}}_i$. Again, for a given initial noise variance in the in-phase quadrature component, $\Delta X_{\Psi}^2(t= 0) \equiv \Delta X_{\text{in}}^2$, the corresponding partition of noise variances in the quadrature components for  two probe fields and atomic polarization operators are:
\begin{eqnarray}
&& \Delta \hat{X}_1^2 = \dfrac{(\dfrac{\Omega_1}{g_1})^2\Delta X_{\text{in}}^2 + (\dfrac{\Omega_2}{g_2})^2 + N}{(\dfrac{\Omega_1}{g_1})^2 + (\dfrac{\Omega_2}{g_2})^2 + N},\\
&&\Delta \hat{X}_2^2 = \dfrac{(\dfrac{\Omega_1}{g_1})^2 + (\dfrac{\Omega_2}{g_2})^2\Delta X_{\text{in}}^2 + N}{(\dfrac{\Omega_1}{g_1})^2 + (\dfrac{\Omega_2}{g_2})^2 + N},\\
&& \Delta \hat{X}_{\sigma}^2 = \dfrac{1}{N}\left[\dfrac{(\dfrac{\Omega_1}{g_1})^2 + (\dfrac{\Omega_2}{g_2})^2 + N \Delta X_{\text{in}}^2}{(\dfrac{\Omega_1}{g_1})^2 + (\dfrac{\Omega_2}{g_2})^2 + N}\right].
\end{eqnarray}
The quantum correlation between each probe field, $\hat{\mathcal{E}}_i$, and the atomic components has the form:
\begin{eqnarray}
\langle\hat{X}_i\hat{X}_{\sigma}\rangle = \dfrac{\Omega_i/g_i}{(\Omega_1/g)^2+(\Omega_2/g)^2+N}[1-\Delta X_\text{in}^2],
\end{eqnarray}
which shares a similar formula as that in a single-$\Lambda$ configuration shown in Eq. (15), except for the addition terms from two prob fields in the denominator. 
The quantum correlation between the quadrature components of two fields is found to have the form:
\begin{eqnarray}
\langle \hat{X}_1\hat{X}_2\rangle= \dfrac{(\dfrac{\Omega_1}{g_1})(\dfrac{\Omega_2}{g_2})}{(\dfrac{\Omega_1}{g_1})^2+(\dfrac{\Omega_2}{g_2})^2+N}[\Delta X_{\text{in}}^2-1].
\label{x1x2}
\end{eqnarray}
In Fig. 5,  we show the quantum correlation between two probe  fields, $\langle\hat{X}_1\hat{X}_2 \rangle$, as a function of the normalized control fields, $\Omega_i/g_i$, $i = 1,2$, for different values of squeezing parameter, $\text{r}$.
When the second probe field is fixed as a constant, for example $\Omega_2/g_2 = 1$, the quantum correlation between two probe-field vanishes as $\Omega_1/g_1 = 0$ or $\Omega_1/g_1 \rightarrow \infty$. Moreover, due to the phase shift, $\pi$, defined for the dark-state polariton in Eq. (20), the correlation between two probe fields in negative (anti-correlated).

\begin{figure}
\includegraphics[width=8.0cm]{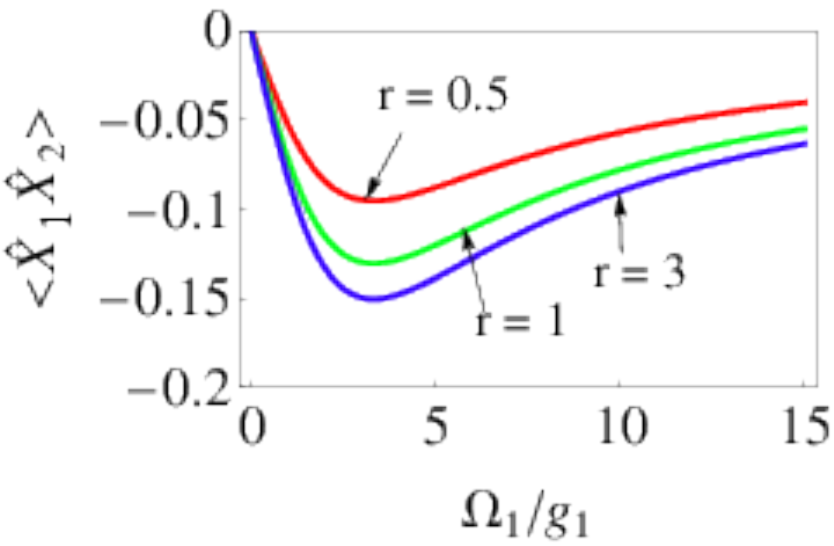} 
\caption{(Color online) Quantum correlation between two probe fields in a double-$\Lambda$ configuration, $\langle \hat{X}_1\hat{X}_2\rangle$, shown as a function of the normalized control field, $\Omega_1/g_1$, for different values of squeezing parameter, $\text{r}$. Here, the other parameters used are  $\Omega_2/g_2=1$ and $N=10$.}
\end{figure}

Below, we show the normalized inseparability criterion for the mutual entanglement among the two probe fields and atomic polarization, denoted as $ \mathcal{I}_c(\mathcal{E}_1, \mathcal{E}_2)$, $\mathcal{I}_c(\mathcal{E}_1,\sigma)$, and $\mathcal{I}_c(\mathcal{E}_2,\sigma)$,  respectively, 
\begin{eqnarray}
\mathcal{I}_c(\mathcal{E}_i,\sigma) &=& 1 + \left( \dfrac{2}{1+1/N}\right) [\dfrac{\mathcal{G}(\Omega_i/g_i,\text{r}, \delta)}{(\dfrac{\Omega_1}{g_1})^2+(\dfrac{\Omega_2}{g_2})^2+N}] \nonumber\\ 
&<&  1, \label{eccc1} \\
\mathcal{I}_c(\mathcal{E}_1,\mathcal{E}_2) &=& 1 + \left( \dfrac{2}{1+1}\right) [\dfrac{\mathcal{H}(\Omega_1/g_1, \Omega_2/g_2, \text{r}, \delta)}{(\dfrac{\Omega_1}{g_1})^2+(\dfrac{\Omega_2}{g_2})^2+N}]\nonumber\\ 
&<&  1, 
\label{eccc2}
\end{eqnarray}
where
\begin{eqnarray}
&&\mathcal{G}(\Omega_i/g_i,\text{r}, \delta)\equiv \label{G}\\
&&\hspace{1.0cm} [(\dfrac{\Omega_i}{g_i})^2+1]\sinh^2 \text{r} - 2\dfrac{\Omega_i}{g_i} \sinh \text{r} \cosh \text{r} \cos\delta, \nonumber\\
&&\mathcal{H}(\Omega_1/g_1, \Omega_2/g_2, \text{r}, \delta)\equiv  \label{H}\\ 
&&\hspace{1.0cm} [(\dfrac{\Omega_1}{g_1})^2+(\dfrac{\Omega_2}{g_2})^2]\sinh^2 \text{r} + 2\dfrac{\Omega_1}{g_1}\dfrac{\Omega_2}{g_2} \sinh \text{r} \cosh \text{r} \cos\delta. \nonumber
\end{eqnarray}
For a given squeezing degree $\text{r}$, we can immediately find the parameter space to  satisfy the inseparability condition to ensure entanglement.

\begin{figure}
\includegraphics[width=16cm]{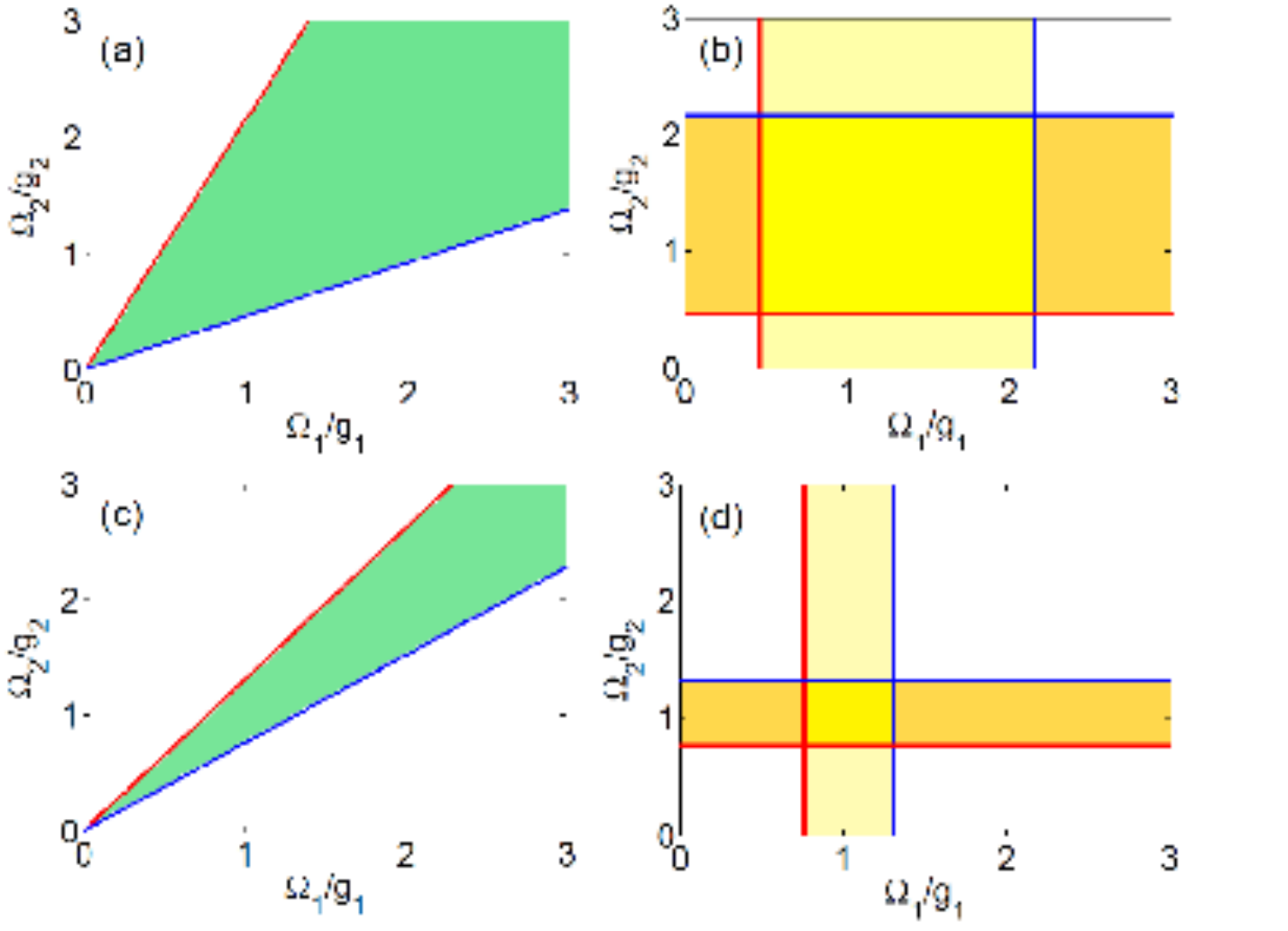} 
\caption{(Color online) Regions to support nonseparated states between field-field and field-atom quadrature components: {\it i.e.}, $\mathcal{I}_c(\mathcal{E}_i,\sigma)$ and $\mathcal{I}_c(\mathcal{E}_1,\mathcal{E}_2)$, as a function of the normalized control fields. 
(a) $\mathcal{I}_c(\mathcal{E}_1,\mathcal{E}_2)< 0$ at $ \text{r}=1$, $\delta=\pi$.
(b) $\mathcal{I}_c(\mathcal{E}_i,\sigma)< 0$ at $\text{r}=1$, $\delta=0$.
(c) $ \mathcal{I}_c(\mathcal{E}_1,\mathcal{E}_2)< 0$ at $\text{r}=2$, $\delta=\pi$.
(d) $ \mathcal{I}_c(\mathcal{E}_i,\sigma)< 0$ at $\text{r}=2$, $\delta=0$. }
\end{figure}

To access these non-classical properties at the output  of atomic ensembles, one can measure the correlations between two probe fields through a homodyne detection. In this scenario, we shown in Fig. 6 the conditions to generate entanglement in the two probe fields, while only one of the input probe fields needs to have non-classical properties. The entanglement is achieved through the collective atoms. 
Moreover,  in Fig. 7, the condition to have entanglement between two probe fields is revealed as a function of  two control fields, $\Omega_1$ and $\Omega_2$. By requiring $\mathcal{H} < 0$ in Eq. (30), we have the following inseparability condition for two probe fields are bounded by two curves:
\begin{eqnarray}
\left( \dfrac{\Omega_2}{g_2}\right) - (A_{-})^{-1} \left( \dfrac{\Omega_1}{g_1}\right) =0,\\
\left( \dfrac{\Omega_2}{g_2}\right) - (A_{+})^{-1} \left( \dfrac{\Omega_1}{g_1}\right) =0,
\end{eqnarray}
with $A_\pm \equiv -\coth(\text{r})\cos\delta\pm \sqrt{\coth^2\text{r}\cos^2\delta - 1}$.
These two curves are plotted in red and blue colors, shown in Fig. 7(a). The colored region within these two curves is the parameter space to have entangled probe fields, {\it i.e.}, $\mathcal{I}_c(\mathcal{E}_1, \mathcal{E}_2) < 0$.
In addition to $\text{r} = 1$, we also show the region to support field-field entanglement for the squeezing parameter $\text{r} = 2$ in Fig. 7(c).
As the same scenario in a single-$\Lambda$ configuration, only a finite range of squeezing parameter supports non-separable states.

Besides direct measurement on the quantum fluctuations in two probe fields in the output, in this double-$\Lambda$ scheme, we can also infer the non-separability between one of the probe fields and collective atomic excitations indirectly. In Fig. 7(b), we demonstrate the entanglement regions for these two probe fields and field-atomic ensembles.
By requiring $\mathcal{G} < 0$ in Eq. (29), the criterion to have entanglement between the output probe field, $\mathcal{E}_1$ or $\mathcal{E}_2$, and the atomic ensemble, {\it i.e.}, $\mathcal{I}_c(\mathcal{E}_i,\sigma) < 0$,  can be achieved when the Rabi frequencies of coupling fields fall in between 
\begin{eqnarray}
B_- < \Omega_i/g_i  < B_+,
\end{eqnarray}
where $B_\pm \equiv \coth(\text{r})\cos\delta\pm \sqrt{\coth^2\text{r}\cos^2\delta - 1}$.
 In this way, one can measure the output probe fields through state-of-the-art quantum detection scheme, which is readily and reliably realized in presently available systems.

However, in terms of the quantized operators, $\hat{\mathcal{E}}_1$, $\hat{\mathcal{E}}_2$, and $\hat{\sigma}$, we can take such a double-$\Lambda$ configuration as a tripartite system.
From the inseparability criterion for field-atom and field-field quadrature components given in Eq. (\ref{eccc1}) and Eq. (\ref{eccc2}), it requires that both $\mathcal{G}(\Omega_i/g_i,\text{r}, \delta)$ and $\mathcal{H}(\Omega_i/g_i,\text{r}, \delta)$ must be negative values simultaneously, in order to have a tripartite entanglement.
It is the phase difference between field and atomic components in the definition of a dark-state polariton shown in Eq. (20),  which automatically results in a $\pi$ phase shift in the squeezing angle. In such a double-$\Lambda$ configuration, it is impossible to support  the co-existence of mutual entanglements among field-field and field-atom simultaneously for this tripartite system.

\section{Conclusion}
In summary, we have introduced the squeezed operator for dark-state polaritons in EIT media, including single- and double-$\Lambda$ configurations.
We show that quantum squeezed state transfer from field to atomic ensemble 
can be achieved by a time-dependent coupling field,  and reveal the quantum correlation and noise entanglement between probe field and atomic polarization.
Even though a larger degree of squeezing parameter in the quadrature  components helps to establish stronger quantum correlations, inseparability criterion is satisfied only within a finite range of squeezing parameter.
The results in our work provide the possible condition to  implement the quantum interface between photon and atomic system.

\section*{Acknowledgment}
We thank Michael Fleischhauer, Gediminas Juzeliunas, Yinchieh Lai, and Yong-Fan Chen for useful discussions. This work was supported by the Frontier Research Center on Fundamental and Applied Sciences of Matters, National Tsing-Hua University, and the Ministry of Science and Technology, Taiwan.


\end{document}